\documentclass[sigconf]{acmart}
\AtBeginDocument{%
  }

\setcopyright{acmlicensed}
\copyrightyear{2026}
\acmYear{2026}
\setcopyright{cc}
\setcctype{by}
\acmConference[WWW '26]{Proceedings of the ACM Web Conference 2026}{April 13--17, 2026}{Dubai, United Arab Emirates}
\acmBooktitle{Proceedings of the ACM Web Conference 2026 (WWW '26), April 13--17, 2026, Dubai, United Arab Emirates}
\acmPrice{}
\acmDOI{10.1145/3774904.3792812}
\acmISBN{979-8-4007-2307-0/2026/04}
\usepackage{amssymb}
\usepackage{booktabs} 

\begin{document}

\title[Latent Reasoning Distillation from Multi-Perspective Chain-of-Thought for E-Commerce Relevance]{\textit{Thinking Broad, Acting Fast}: Latent Reasoning Distillation from Multi-Perspective Chain-of-Thought for E-Commerce Relevance}

\author{Baopu Qiu}
\authornote{Both authors contributed equally to this research.}
\email{qiubaopu.qbp@alibaba-inc.com}
\orcid{0009-0003-6872-4312}
\author{Hao Chen}
\authornotemark[1]
\email{ryan.ch@alibaba-inc.com}
\orcid{0009-0008-9029-8571}
\affiliation{%
  \institution{Alibaba International Digital Commerce Group}
  \city{Hangzhou}
  \country{China}
}

\author{Yuanrong Wu}
\orcid{0009-0003-0404-7594}
\authornote{Work done during internship at Alibaba International Digital Commerce Group.}

\email{wuyr23@zju.edu.cn}
\affiliation{%
  \institution{Zhejiang University}
  \city{Hangzhou}
  \country{China}
}

\author{Changtong Zan}
\orcid{0000-0002-5467-0937}
\email{zanchangtong.zct@alibaba-inc.com}
\affiliation{%
  \institution{Alibaba International Digital Commerce Group}
  \city{Hangzhou}
  \country{China}
}

\author{Chao Wei}
\orcid{0009-0006-2417-6029}
\authornote{Corresponding author.}
\email{weichao.wc@alibaba-inc.com}
\affiliation{%
  \institution{Alibaba International Digital Commerce Group}
  \city{Hangzhou}
  \country{China}
}

\author{Weiru Zhang}
\orcid{0009-0007-8496-3638}
\email{weiru.zwr@alibaba-inc.com}
\affiliation{%
  \institution{Alibaba International Digital Commerce Group}
  \city{Hangzhou}
  \country{China}
}

\author{Xiaoyi Zeng}
\orcid{0000-0002-3742-4910}
\email{yuanhan@taobao.com}
\affiliation{%
  \institution{Alibaba International Digital Commerce Group}
  \city{Hangzhou}
  \country{China}
}

\renewcommand{\shortauthors}{Baopu Qiu et al.}

\begin{abstract}
Effective relevance modeling is crucial for e-commerce search, as it aligns search results with user intent and enhances customer experience. Recent work has leveraged large language models (LLMs) to address the limitations of traditional relevance models, particularly their inability to handle long-tail and ambiguous queries. By incorporating Chain-of-Thought (CoT) reasoning, these approaches further improve both accuracy and interpretability through explicit, multi-step reasoning pathways. However, two key limitations remain: (1) most existing approaches rely on single-perspective CoT reasoning, which fails to capture the multifaceted nature of e-commerce relevance (e.g., user intent vs. attribute-level matching vs. business-specific rules); and (2) although CoT-enhanced LLMs offer rich reasoning capabilities, their high inference latency necessitates knowledge distillation for real-time deployment, yet current distillation methods discard the CoT rationale structure at inference, using it only as a transient auxiliary signal and thereby forfeiting its reasoning utility for online serving. To address these challenges, we propose a novel framework that better exploits CoT semantics throughout the optimization pipeline. Specifically, the teacher model leverages Multi-Perspective CoT (MPCoT) to generate diverse rationales and combines Supervised Fine-Tuning (SFT) with Direct Preference Optimization (DPO) to construct a more robust reasoner. For distillation, we introduce Latent Reasoning Knowledge Distillation (LRKD), which endows a student model with a lightweight inference-time latent reasoning extractor, allowing efficient and low-latency internalization of the LLM’s sophisticated reasoning capabilities. Evaluated through offline experiments and online A/B tests on an e-commerce search advertising platform serving tens of millions of users daily, our method delivers significant offline gains, along with a 1.42\% online improvement in Revenue Per Mille (RPM) and a 0.4\% increase in relevance satisfaction score (RS), demonstrating clear benefits in both commercial performance and user experience.

\end{abstract}




\maketitle

\section{Introduction}
Relevance classification in e-commerce generally refers to the task of estimating how well an item product satisfies a user's query, which is fundamental to deliver accurate search results, enhance user experience, and drive business outcomes. In large-scale platforms such as Amazon\footnote{\url{https://www.amazon.com}} and AliExpress\footnote{\url{https://www.aliexpress.com}}, this task must simultaneously achieve high accuracy and computational efficiency to handle massive daily traffic and thousands of candidate items per query. The field has evolved significantly, progressing from early statistical methods~\citep{ramos2003using,robertson2009probabilistic} to neural approaches~\citep{10.1145/2505515.2505665,Palangi2014SemanticMW,kenton2019bert,khattab2020colbert,reimers2019sentence,humeaupoly} that better capture deep semantic matching. However, while these models effectively handle 80–90\% of user queries, the remaining 10\%+ comprising long-tail and ambiguous cases are still challenging, as they require more sophisticated reasoning. Critically, these hard cases often correspond to high-intent user journeys, where accurate understanding directly determines satisfaction and trust. Therefore, reliably resolving them is not only vital for user experience but also a key differentiator among competitive e-commerce platforms. To address these limitations, recent work has explored large language models (LLMs) for relevance classification~\citep{10.1145/3578337.3605136,yan2024consolidating,mehrdad2024large,wang2025llm}, showing substantial gains via prompting or fine-tuning. Nevertheless,  the high latency and computational cost of LLMs render them impractical for real-time serving, making knowledge distillation a standard approach to transfer their capabilities to efficient student models~\citep{Zhao_2025}. Recent efforts further distill LLM-generated rationales into student models to preserve reasoning semantics~\citep{agrawal-etal-2025-rationale,Zhao_2025}.

\begin{figure}
    \centering
    \includegraphics[width=\linewidth]{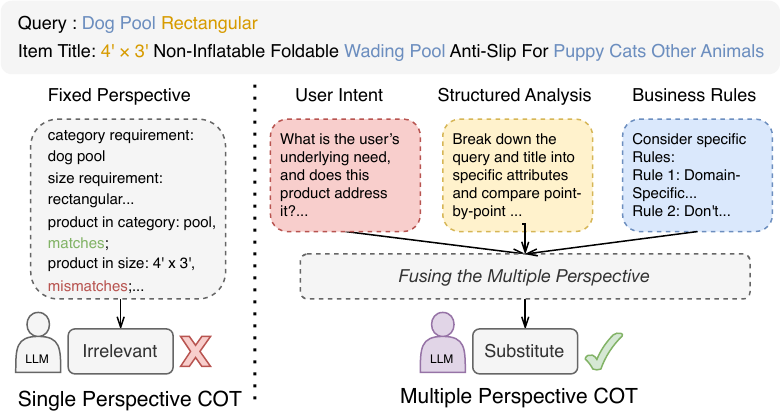}
    \caption{From Single to Multiple Perspectives in Reasoning}
    \Description{Illustration showing the transition from single to multiple perspectives in reasoning.}
    \label{fig:mp_cot_demo}
\end{figure}

However, existing approaches suffer from two key limitations in the context of e-commerce relevance modeling. 
First, they often rely on \textit{single-perspective reasoning}, where the LLM generates relevance judgments from a single, monolithic viewpoint and ignores the multifaceted nature of e-commerce relevance. In practice, different perspectives yield complementary insights: for instance, a \textit{User Intent} perspective focuses on functional needs and use-case alignment (e.g., ``Will this product serve my purpose?''), while an \textit{Structured Analysis} perspective systematically verifies point-wise matches between query specifications and product attributes (e.g., size, color, category), and a \textit{Business Rule} perspective enforces platform-specific heuristics to resolve ambiguities (e.g., distinguishing main products from accessories). As illustrated in Figure~\ref{fig:mp_cot_demo}, these diverse reasoning paths often lead to divergent conclusions, and fusing them adaptively is crucial for robust relevance judgment. Second, current distillation methods that leverage LLM-generated rationales typically discard the rationale structure at inference time, which reduces it to a transient training signal and limits their online utility in the deployed student model.


To address these gaps, we propose a framework that \textit{thinks broad} during training and \textit{acts fast} at inference. Our main contributions can be summarized as follows:
\begin{enumerate}
    \item We introduce Multi-Perspective Chain-of-Thought (MPCoT), a novel reasoning framework that generates diverse and complementary rationales from distinct perspective to capture the multifaceted nature of e-commerce relevance more comprehensively than single-perspective CoT approaches.

    \item We propose Latent Reasoning Knowledge Distillation (LRKD), which distills the reasoning semantics of MPCoT-enhanced LLMs into a lightweight student model (e.g., BERT) via a trainable latent reasoning extractor. Unlike prior methods that discard rationales at inference, LRKD retains reasoning capabilities in a compact, non-generative form by aligning latent representations with CoT embeddings.
    
    \item Extensive offline experiments on multilingual e-commerce datasets (AliExpress and ESCI) demonstrate that the combination of MPCoT and LRKD consistently outperforms strong baselines, with ablation and probing studies confirming that the extractor captures high-level reasoning beyond surface lexical cues. We further validate real-world impact through a large-scale online A/B test on AliExpress’s search advertising platform, achieving statistically significant gains of \textbf{+1.42\%} in Revenue Per Mille (RPM), \textbf{+0.48\%} in Click-Through Rate(CTR), and \textbf{+0.4\%} in relevance satisfaction (RS) score.
\end{enumerate}

\section{Related Works}
\subsection{Relevance Modeling for E-Commerce}
Relevance modeling is a core component of e-commerce search, which aims to estimate the semantic match between user queries and items. Early approaches relied on statistical methods like TF-IDF~\citep{ramos2003using} and BM25~\citep{robertson2009probabilistic}, which are efficient but limited in capturing deep semantic relationships. With the development of deep learning, neural methods have been applied in the relevance modeling task. The first wave involves learning dense representations for queries and items using the dual-encoder model \cite{10.1145/2505515.2505665,Palangi2014SemanticMW}, but their potential is constrained by the weak interaction between the query and item representations. This limitation was addressed by interaction-based models such as BERT\cite{kenton2019bert}, Colbert\citep{khattab2020colbert} and PolyEncoder \citep{humeaupoly}, which explore token-level interaction through cross-attention or dedicated interaction modules, yielding strong relevance performance in e-commerce search. Despite their success, bert-style models remain constrained on long-tail or ambiguous queries, motivating recent efforts to leverage Large Language Models (LLMs) for more robust and reasoning-aware relevance judgment.

\begin{figure*}
    \centering
    \includegraphics[width=\linewidth]{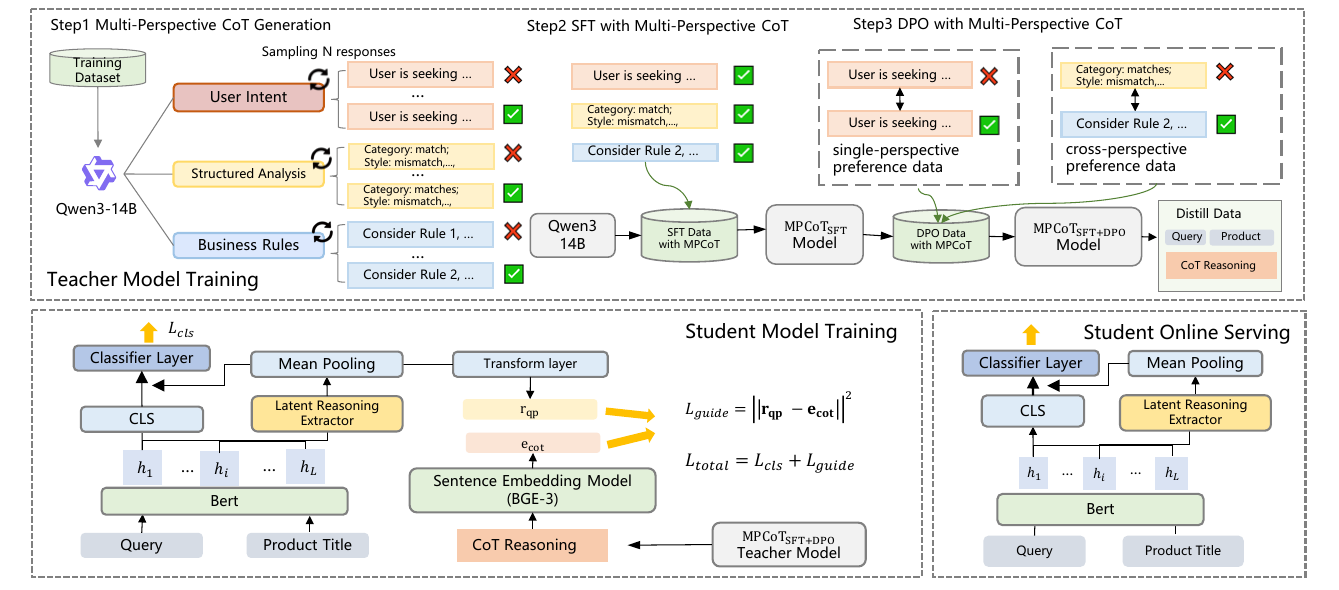}
    \caption{Framework overview of MPCoT teacher training (top) and LRKD student distillation and deployment (bottom).}
    \Description{Top: Multi-Perspective CoT (MPCoT) teacher training with SFT and cross-perspective DPO. Bottom: Student model with latent reasoning extractor—used during distillation (with CoT guidance loss) and retained for efficient online serving.}
    \label{fig:overall}
\end{figure*}

\subsection{LLM for Relevance}
Large Language Models (LLMs) have shown strong performance for e-commerce relevance via prompting or fine-tuning in recent works ~\cite{10.1145/3578337.3605136,yan2024consolidating,mehrdad2024large,wang2025llm}. However, treating LLMs as black-box annotators limits transparency and hinders advanced optimization (e.g., reinforcement learning~\cite{schulman2017proximal,rafailov2023direct}). To improve interpretability, recent approaches elicit Chain-of-Thought (CoT) alongside predictions~\cite{wei2022chain}. For example, \citet{Zhao_2025} models relevance as an interpretable CoT reasoning process and distills both rationales and score distributions from an explainable LLM to improve student models’ semantic interaction, while \citet{Tang_2025} employs a multi-stage SFT with diverse CoT strategies to guide the model in their LREF framework. Other works improve LLM reasoning via external signals or reinforcement learning, including incorporating user behavior~\cite{chen2025towards}, applying DPO to reduce optimistic bias~\cite{rafailov2023direct,Tang_2025}, or using GRPO to mitigate reasoning hallucinations~\cite{shao2024deepseekmath,dong_taosr1_2025}.  

\subsection{Knowledge Distillation for Relevance}
Knowledge Distillation (KD)~\cite{hinton2015distilling} is widely used to deploy large models in real-world search systems. Early work focused on compressing Transformer-based rankers (e.g., BERT) via score-based distillation~\cite{liu2022knowledge,reddi2021rankdistil,zhuang2021ensemble}, often using specialized losses like Margin-MSE~\cite{hofstatter2020improving}. With the rise of LLMs as powerful teachers, recent methods distill LLM-generated pseudo-labels into smaller models like BERT~\cite{ye2025best,shang2025knowledge,ye2025applying}. However, most treat the LLM as a black-box annotator, underutilizing its rich reasoning. A few recent works begin to distill Chain-of-Thoughts (CoTs): \citet{Zhao_2025} use token-level tagging to transfer attribute-level reasoning, while \citet{agrawal-etal-2025-rationale} train a decoder to reconstruct rationales as an auxiliary task. However, these methods discard rationales at inference, limiting the utility of the online environment.

\section{Methods}
As shown in Figure~\ref{fig:overall}, our method consists of two components: Multi-Perspective Chain-of-Thought (MPCoT) for teacher reasoning and Latent Reasoning Knowledge Distillation (LRKD) for student training. We first define the e-commerce relevance task, then motivate multi-perspective reasoning with empirical analysis, and detail how we optimize the teacher via SFT and DPO. Finally, we present LRKD in Section~\ref{sec:LRKD}, which distills the teacher’s reasoning into a lightweight student model.

\subsection{Relevance Task Definition}
In e-commerce search, relevance modeling aims to estimate the semantic match between a user's query $q$ and a product item title $p$, represented as a pair $(q, p)$. We formulate this task as a \textbf{multi-class classification problem}, where the goal is to predict a relevance label $y$ from a predefined set of $C$ discrete classes, $\mathcal{Y} = \{c_1, c_2, \dots, c_C\}$. Rather than adopting a simple binary schema (e.g., \{\textit{Good}, \textit{Bad}\}), we use a fine-grained multi-class label space (see \ref{sec:dataset} for details) for two reasons: (1) it poses a more challenging and realistic modeling task for precise judgment of relevance states, and (2) it provides richer supervision signals that allow large language models to leverage their advanced reasoning and discrimination capabilities. The exact formulation of the task differs for our student and teacher models, reflecting their distinct capabilities:

\textbf{Student Model: Direct Classification.} For the lightweight student model (e.g., BERT), the task is to learn a direct mapping function $f_{\text{student}}: (q, p) \to \mathbf{z}$, where $\mathbf{z} \in \mathbb{R}^C$ is the logit vector over the $C$ classes. The final probability distribution is obtained via the softmax function, $P(y|q,p) = \text{softmax}(\mathbf{z})$.

\textbf{Teacher Model: CoT-based Reasoning.} In contrast, for the LLM-based teacher model, we reformulate the task to explicitly generate a reasoning process. The model learns a conditional generation function $f_{\text{teacher}}:(q, p)\to(r, y)$, where $r$ is a natural language rationale (Chain-of-Thought) that justifies the prediction, and $y \in \mathcal{Y}$ is the final relevance label. This transforms the task from a simple classification into an explainable reasoning problem.



\subsection{Multi-Perspective Chain-of-Thought}

\begin{figure}
    \centering
    \includegraphics[width=\linewidth]{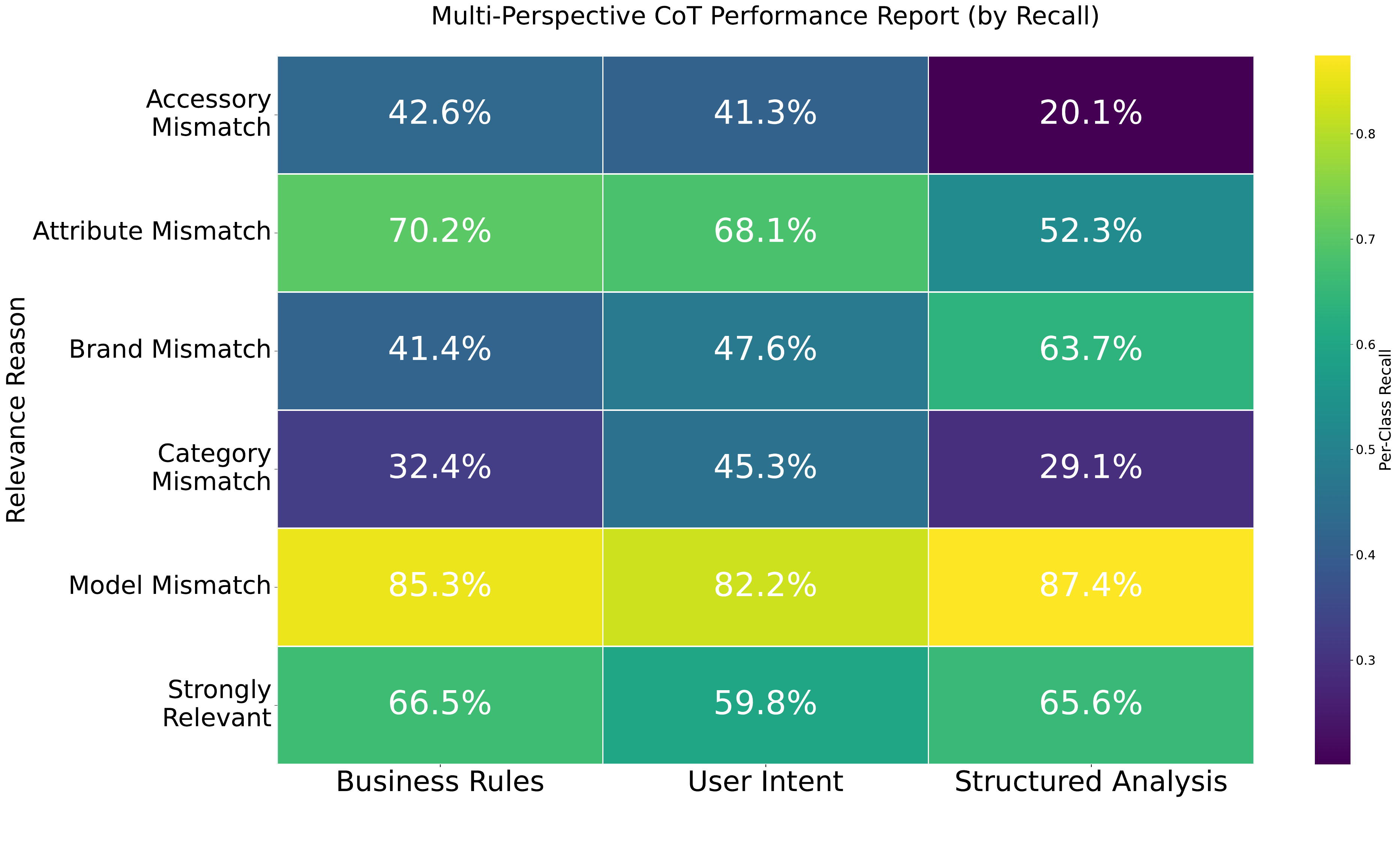}
    \caption{Performance comparison of Chain-of-Thought (CoT) generated from different perspectives on the AliExpress dataset.}
    \Description{Performance comparison heatmap of Chain-of-Thought (CoT) generated from different perspectives on the AliExpress dataset.}
    \label{fig:MPCoT_analysis}
\end{figure}


The foundation of our framework is a powerful teacher model capable of deep reasoning, so that it can distill comprehensive supervisory signals to a lightweight student model . While existing LLM-based relevance approaches have leveraged techniques like Chain-of-Thought (CoT) \cite{Zhao_2025} to enhance explainability or Retrieval-Augmented Generation (RAG) \cite{dong_taosr1_2025} to incorporate more product information, their reliance on a single, monolithic reasoning perspective limits the depth and breadth of the knowledge they can provide, as e-commerce platforms involves a complex interplay of diverse user intents and pragmatic business rules. 

To empirically validate our multi-perspective hypothesis, we conducted a pilot experiment by directly prompting a base LLM (e.g., Qwen3-14B) to generate a relevance judgment along with a rationale. This was done independently for three perspectives:
\begin{itemize}
    \item \textbf{User Intent Perspective.} Simulates a shopper's intuitive thought process, focusing on use-case scenarios and functional needs.

    \item \textbf{Structured Analysis Perspective.} Mimics an expert annotator by performing a systematic, attribute-by-attribute analysis of the product against the query. 
    
    \item \textbf{Business Rule Perspective.} Applies a set of heuristics to resolve common ambiguous cases in e-commerce, such as distinguishing a main product from its accessories.
\end{itemize}
As shown in Figure~\ref{fig:MPCoT_analysis}, our analysis of the in-house AliExpress dataset reveals that each perspective exhibits distinct and complementary strengths in handling different relevance issues. For instance, the \textit{User Intent} perspective is highly effective at identifying \textit{Category Mismatch} issues, reflecting a user's high sensitivity to whether a product belongs to the correct category. Conversely, the \textit{Structured Analysis} perspective excels at identifying \textit{Model Mismatch} and \textit{Brand Mismatch} through systematic checks, yet it interestingly struggles with fine-grained \textit{Attribute Mismatch}. This struggle arises because judging such cases often requires specific business rules. The \textit{Business Rule} perspective is better suited for this and, in turn, effectively resolves Accessory Mismatch ambiguities.

To isolate the benefit of perspective diversity from merely increasing the number of generation attempts, we conducted a controlled experiment. First, we established a multi-perspective oracle by prompting the base LLM (Qwen3-14B) independently in few-shot setting with each of the three perspectives; an input was considered correctly classified if any of the three outputs was correct. This oracle achieved an accuracy of 77.52\%. For a fair comparison, we created a strong control baseline: a single-perspective pass@3, which also generates three outputs but uses the same top-performing single perspective (\textit{User Intent}) for all attempts. The multi-perspective oracle significantly outperforms not only the single-pass baseline (\textit{User Intent}, 64.61\%) but also, more critically, the pass@3 control (72.59\%). This result provides strong evidence that the performance gain is attributable to genuine perspective diversity, not simply an increased generation budget.



\subsubsection{SFT with Multi-Perspective CoT}

Our approach to Supervised Fine-Tuning (SFT) is designed to endow the teacher model with robust, multi-faceted reasoning capabilities. Unlike methods that rely on extensive data filtering or knowledge distillation from much larger models, we introduce a \textbf{self-enhancement training loop} that leverages the model's own reasoning abilities, enriched by our multi-perspective framework. The process consists of two stages: Multi-Perspective CoT Data Generation and Fine-tuning.

\paragraph{Multi-Perspective CoT Data Generation.}
We begin with our human-annotated seed dataset $\mathcal{D} = \{(q_i, p_i, y_i)\}_{i=1}^N$. For each sample $(q, p)$, we prompt the base model $\mathcal{M}$ to generate a CoT-augmented output $(r_k, \hat{y}_k)$ for each of the three perspectives $k \in \{\text{User, Struct, Rule}\}$. To ensure the quality of these generated rationales, we apply a \textbf{consistency-based filtering} step. We only retain outputs where the model's predicted label $\hat{y}_k$ matches the ground-truth label $y$:
\begin{equation}
\label{eq:sft_data_gen}
\mathcal{D}_k^{\text{CoT}} = \{ (q_i, p_i, r_{ik}, y_i) \mid (r_{ik}, \hat{y}_{ik}) = \mathcal{M}(q_i, p_i, \text{prompt}_k) \land \hat{y}_{ik} = y_i \}
\end{equation}
This results in three high-quality, perspective-specific datasets.

\paragraph{Fine-tuning on Aggregated Knowledge.}
We then create a training dataset $\mathcal{D}_{\text{SFT}}$ by aggregating all the perspective-specific data:
\begin{equation}
\label{eq:sft_data_agg}
\mathcal{D}_{\text{SFT}} = \bigcup_{k \in \{\text{User, Struct., Rule}\}} \mathcal{D}_k^{\text{CoT}}
\end{equation}
The model is fine-tuned on this aggregated dataset using a standard "think-then-respond" format. A unified instruction prompt is used for all samples, which instruct the model to generate the correct rationale and label $(r, y)$ without specifying which perspective to adopt. The objective is to minimize the negative log-likelihood:
\begin{equation}
\label{eq:sft_loss}
\mathcal{L}_{\text{SFT}} = - \sum_{(q, p, r, y) \in \mathcal{D}_{\text{SFT}}} \log P(r, y \mid q, p; \theta)
\end{equation}
This process trains the model on a diverse corpus of valid reasoning paths, forcing it to internalize the nuances of different relevance criteria and thereby enhancing its adaptability.

\subsubsection{DPO with Multi-Perspective CoT}

While the SFT model trained with multi-perspective CoT data possesses diverse reasoning capabilities, it does not explicitly learn how to choose the optimal perspective for a given query-product context. To address this, we introduce a DPO strategy that leverages our multi-perspective framework to refine the model's preference alignment on challenging cases, particularly those that result in conflicting predictions among the perspectives.

Our approach first identifies samples with conflicting predictions, defined as any sample misclassified by at least one of the three perspectives. This set, $\mathcal{D}_{\text{conflict}}$, is the union of erroneously predicted samples from each perspective $k \in \{\text{User, Struct, Rules}\}$:
\begin{equation}
\label{eq:dpo_conflict_set}
\mathcal{D}_{\text{conflict}} = \{ (q, p, y) \mid \exists k, \mathcal{M}_{\text{SFT}}(q, p, \text{prompt}_k) \rightarrow \hat{y}_k \neq y \}
\end{equation}
For each sample in this set, we employ a \textbf{cross-perspective preference construction} method. We designate an incorrect rationale from one of the failing perspectives as the "rejected" response ($y^-$). The "chosen" response ($y^+$) is then sourced from a correct rationale found within multiple generation attempts (\texttt{pass@5}) across all three perspectives. This strategy is particularly effective for guiding the model on how to weigh conflicting signals. For example, consider a sample misclassified by the \textit{User Intent} perspective but correctly classified by the \textit{Structured Analysis} perspective. By pairing the incorrect \textit{User Intent} rationale as $y^-$ and the correct attribute-based \textit{Structured Analysis} rationale as $y^+$, the model learns to prioritize the latter in this context.

We construct the full preference dataset $\mathcal{D}_{\text{DPO}}$ in this manner. The model is then fine-tuned by minimizing the standard DPO loss:
\begin{equation}
\label{eq:dpo_loss}
\mathcal{L}_{\text{DPO}} = - \mathbb{E}_{(q,p,y^+,y^-) \sim \mathcal{D}_{\text{DPO}}} \left[ \log \sigma \left( \pi_{\theta}(y^+ | q,p) - \pi_{\theta}(y^- | q,p) \right) \right]
\end{equation}
This DPO process guides the model to learn the subtle, context-dependent preferences for different reasoning styles. This effectively trains the model to better reconcile its own diverse viewpoints when making a final judgment.

\subsection{Latent Reasoning Knowledge Distillation}
\label{sec:LRKD}
We propose Latent Reasoning Knowledge Distillation (LRKD), a framework that transfers the semantic content of LLM-generated Chain-of-Thought (CoT) into lightweight cross-encoders for fine-grained e-commerce relevance classification. Unlike binary score-based distillation, which relies on $p_{\text{Good}}/(p_{\text{Good}} + p_{\text{Bad}})$ ratios and is inapplicable to our multi-class setting, LRKD distills full rationale semantics without requiring text generation during training or inference.


\subsubsection{Model Structure}
Our student model consists of two components, as in Figure \ref{fig:overall}:(1) A BERT-based cross encoder that encodes the concatenated input $[q; p]$ into a sequence of contextualized token representations $\mathbf{H} \in \mathbb{R}^{L \times d}$, where $L$ is the sequence length and $d$ is the hidden dimension; 
(2) A trainable latent reasoning extractor $\mathcal{R}(\cdot)$ maps $\mathbf{H}$ to a structured latent reasoning vector:
\begin{equation}
    \mathbf{r}_{qp} = \mathcal{R}(\mathbf{H}, \mathbf{M}_{qp}) \in \mathbb{R}^d,
\end{equation}
where $\mathbf{M}_{qp}$ is the attention mask for the query-item title sequence.
Separately, we use a frozen high-performance sentence embedding model (e.g., BGE-M3) to encode an LLM-generated chain of thought (CoT) reasoning $r$ for the same pair $(q, p)$: $\mathbf{e}_{\text{cot}} = \text{BGE}(r) \in \mathbb{R}^d$, which serves as a semantic guidance signal that captures the LLM’s reasoning intent in a dense vector space. We explore several lightweight extractor architectures, including MLP, Graph Attention Network (GAT)~\cite{velickovic2017graph} and Poly-Encoder~\cite{humeaupoly}, all of which operate on contextualized token representations from the BERT encoder. These designs vary in how they aggregate token-level signals into a structured latent reasoning vector, and we empirically compare their effectiveness in Section~\ref{sec:experiments}. Full mathematical formulations are provided in the Appendix~\ref{app:extractor_details}.


\subsubsection{Training Objective}
The training objective combines two losses:
\begin{enumerate}
    \item \textbf{Multi-Class Relevance Classification Loss.} A standard categorical cross-entropy loss over $C$ relevance classes:
    \begin{equation}
        \mathcal{L}_{\text{cls}} = -\frac{1}{N} \sum_{i=1}^{N} \sum_{c=1}^{C} \mathbb{1}(y_i = c) \cdot \log \left( \frac{\exp(z_{i,c})}{\sum_{c'=1}^{C} \exp(z_{i,c'})} \right),
    \end{equation}
    where $ \mathbb{1}(\cdot) $ is the indicator function, and final logits $ \mathbf{z}_i \in \mathbb{R}^C $ are obtained by projecting $\mathbf{h}_{\text{fuse}} $, which concatenates the BERT [CLS] representation $\mathbf{h}_{\text{cls}}$ and the latent reasoning vector $ \mathbf{r}_{qp} $ through a linear classification head.
    
    \item \textbf{Latent Reasoning Guidance Loss.} A mean squared error (MSE) loss that guides the student’s latent reasoning representation toward the teacher’s CoT embedding:
    \begin{equation}
        \mathcal{L}_\text{guide} = \|\mathbf{r}_{qp} - \mathbf{e}_{\text{cot}}\|_2^2.
    \end{equation}
\end{enumerate}

The full training objective is a weighted sum:
\begin{equation}
    \mathcal{L}_{\text{total}} = \mathcal{L}_{\text{cls}} + \lambda \mathcal{L}_{\text{guide}},
\end{equation}
where $\lambda \geq 0$ is a hyperparameter that controls the strength of latent reasoning guidance.
LRKD retains the latent reasoning extractor at inference time, which directly contributes to final predictions. 


\section{Experiments}
\label{sec:experiments}

\subsection{Dataset}
\label{sec:dataset}
We conducted experiments on public and in-house datasets respectively, which were collected from real-world e-commerce platforms.
\begin{itemize}
    \item \textbf{Amazon Shopping Queries ESCI Dataset\cite{reddy2022shoppingqueriesdatasetlargescale}.}\footnote{\url{https://github.com/amazon-science/esci-data}} This dataset contains manually labeled query–product relevance judgments from the Amazon online marketplace in three languages: English (US), Spanish (ES) and Japanese (JP), which introduces a fine-grained four-class relevance schema: \{\textit{Exact(E)}, \textit{Substitute(S)}, \textit{Complement(C)}, \textit{Irrelevant(I)}\}. We sample 100K training and 10K test samples from the large version dataset for SFT and DPO training, as the same setting \citet{agrawal-etal-2025-rationale} mentioned.

    \item \textbf{AliExpress Dataset.} This multilingual dataset is collected from AliExpress, a cross-border e-commerce platform operated by Alibaba. To compare with the ESCI dataset at similar magnitude, we sample approximately 100K query–product pairs for training and 20K for testing. The dataset covers six major languages: English (EN), Spanish (ES), Korean (KO), Japanese (JP), Portuguese (PT), and French (FR), and uses a six-class fine-grained relevance schema: \{\textit{Strongly Relevant}, \textit{Brand Mismatch}, \textit{Category Mismatch}, \textit{Model Mismatch}, \textit{Attribute Mismatch}, \textit{Accessory Mismatch}\}. Note this sampled distribution does not reflect actual online traffic patterns.
\end{itemize}
Both datasets are used consistently throughout the SFT, DPO, and knowledge distillation stages. For a fair comparison between teacher and student models, we employ the same training and test splits during distillation as those used for teacher model training.

\begin{table*}[htbp]
\centering
\caption{Comparison of different methods on AliExpress and ESCI test sets. }
\label{tab:main_results}
\begin{tabular}{lcccccccc}
\toprule
\textbf{Model} &
\multicolumn{2}{c}{\textbf{AliExpress}} &
\multicolumn{2}{c}{\textbf{ESCI-US}} &
\multicolumn{2}{c}{\textbf{ESCI-JP}} &
\multicolumn{2}{c}{\textbf{ESCI-ES}} \\
\cmidrule(lr){2-3} \cmidrule(lr){4-5}\cmidrule(lr){6-7} \cmidrule(lr){8-9}
& \textbf{ACC} & \textbf{F1} & \textbf{ACC} & \textbf{F1} & \textbf{ACC} & \textbf{F1} & \textbf{ACC} & \textbf{F1} \\
\midrule
\multicolumn{9}{l}{\textbf{Teacher Model:}} \\
\quad $\text{Best-SingleCoT}_{\text{few-shot}}$ & 54.92 & 47.01 & 59.39 & 48.05 &58.72  & 48.90 & 58.95 & 52.88 \\
\quad $\text{Best-SingleCoT}_{\text{SFT}}$  & 59.81 & 53.87 & 63.20 & \underline{50.21} & 60.11 &49.97  & 60.62 & 55.28 \\
\quad $\text{Best-SingleCoT}_{\text{SFT+DPO}}$ & \underline{64.83} & 58.26 & \underline{69.49} & 50.08  & \underline{65.39} & \underline{51.14} & \textbf{68.10} & \underline{58.21} \\
\quad $\text{ProgressiveCoT}_{\text{SFT}}$\cite{Tang_2025} & 61.47 & 54.22 & 62.86 & 48.45 & 62.33 & 50.01 & 61.03 & 54.05 \\
\quad $\text{MPCoT}_{\text{few-shot}}$  & 55.86 & 48.00 & 57.51 & 45.09 & 55.82 & 45.61 & 57.62 & 49.49 \\
\quad $\text{MPCoT}_{\text{SFT}}$ & 64.45 & \underline{62.18} & 64.26 & 49.48 & 64.17 & 48.06 & 61.37 & 56.02\\
\quad $\text{MPCoT}_{\text{SFT+DPO}}$ & \textbf{68.36} & \textbf{65.90} & \textbf{70.23} & \textbf{51.85} & \textbf{69.72} & \textbf{53.69} & \underline{67.32} & \textbf{60.31}\\
\multicolumn{9}{l}{\textbf{Deployable Model Baseline:}} \\
\quad BERT-multilingual-base \cite{kenton2019bert} & 53.17 & 49.34 & 66.83 & 41.38 & 62.36 & 42.60 & 64.50 & 53.25 \\
\multicolumn{9}{l}{\textbf{Student Model:}} \\
\quad CED-KD \cite{agrawal-etal-2025-rationale} & 56.82 & 50.87 & 67.49  & 43.71  & 62.94 & 43.10  & 64.51 & 53.61  \\
\quad MKD \cite{Zhao_2025} & 55.93 & 50.65  & \underline{68.34} & 42.45 & 62.50 & 43.17 & 65.64 & 54.93 \\
\addlinespace
\quad $\text{LRKD}_{\text{MLP}}$ & 56.04 & 50.98 & 66.94 & 42.61 & 62.19 & 42.96   & 64.99 & 54.61  \\
\quad $\text{LRKD}_{\text{Poly}}$ & \underline{56.92} & \underline{51.37} & 67.92 & \underline{44.33}  & \underline{63.44} &  \underline{45.27} & \underline{65.89} & \underline{55.47}  \\
\quad $\text{LRKD}_{\text{GAT}}$ & \textbf{57.36}  & \textbf{52.29} & \textbf{68.73} &\textbf{45.83}  & \textbf{63.96}  & \textbf{47.21} & \textbf{66.30} & \textbf{55.89}  \\
\bottomrule
\end{tabular}
\end{table*}

\subsection{Baseline and Evaluation Metrics}
\subsubsection{Baselines}
We compare our framework against a set of representative baselines across different stages of the LLM-to-deployable-model pipeline:
During the teacher training stage, we compare our multi-perspective teacher model against its strongest single-perspective counterparts, which are established by selecting the \textit{top-performing} model among the three perspectives (User, Struct., and Rules) at each training phase: \textbf{(1)~$\text{Best-SingleCoT}_{\text{few-shot}}$}, the top performer using few-shot prompting; \textbf{(2)~$\text{Best-SingleCoT}_{\text{SFT}}$}, the top performer after the SFT stage; and \textbf{(3)~$\text{Best-SingleCoT}_{\text{SFT+DPO}}$}, the top performer after the full SFT+DPO pipeline, where the DPO stage is trained on preference pairs constructed from 5 generation passes. For a broader comparison, we also include \textbf{(4)~an external baseline, $\text{ProgressiveCoT}_{\text{SFT}}$}, inspired by the curriculum-based SFT approach of \citet{Tang_2025}.

During the distillation stage, We compare against three representative baselines: (1)~\textit{BERT-multilingual-base}\cite{kenton2019bert}, a standard multilingual BERT model fine-tuned on human-annotated query–item pairs for relevance classification without any rationale distillation; (2)~\textit{Cross-Encoder-Decoder Knowledge Distillation (CED-KD)}~\cite{agrawal-etal-2025-rationale}, which trains a BERT cross-encoder alongside an auxiliary decoder to reconstruct LLM-generated rationales during training—the decoder is discarded at inference; and (3)~\textit{Multi-Dimensional Knowledge Distillation (MKD)}~\cite{Zhao_2025}, which extracts the LLM’s CoTs into token-level BIO tags and guides the student model to recover them via a CRF-based distillation loss.


\subsubsection{Evaluation Metrics}
For offline evaluation,  we adopt Accuracy (ACC) and macro-F1 score (F1) as our core metrics, which are widely recognized in e-commerce relevance evaluation. For online evaluation, we deploy our model in a large-scale A/B test on AliExpress’s search advertising platform. Key online metrics include: (1) Revenue Per Mille (RPM), defined as the revenue generated per thousand ad impressions; (2) Click-Through Rate (CTR), the ratio of ad clicks to impressions; and (3) Relevance Satisfaction Score (RS), a human-evaluated metric based on expert annotations of randomly sampled query–ad item pairs from live traffic.

\subsection{Implementation Details}
All teacher models are based on Qwen3-14B\cite{yang2025qwen3} with LoRA for SFT (3 epochs, lr=$5\!\times\!10^{-5}$, batch=32) and DPO (3 epochs, lr=$5\!\times\!10^{-6}$, batch=64). All student models use BERT-multilingual-base\cite{kenton2019bert} as the backbone and trained for 5 epochs (lr=$2\!\times\!10^{-6}$, batch=256) with 128-token input. In LRKD, the guidance loss weight is set to $\lambda=0.1$. Full details for reproducibility are provided in the Appendix~\ref{app:implementation_details}.

\subsection{Performance}

\paragraph{Overall Performance}
As shown in Table~\ref{tab:main_results}, our complete framework, encompassing the MPCoT teacher and the LRKD student, achieves state-of-the-art results across nearly all datasets. The multi-perspective-based SFT and DPO forge a powerful teacher model that learns to adaptively select the optimal reasoning path based on the context. This expert model is then effectively distilled into a compact student using our LRKD framework. In particular, our stronger LRKD variants including GAT and Poly-encoder consistently outperform previous state-of-the-art distillation methods, demonstrating the framework's superior ability to distill the teacher's sophisticated and multi-faceted relevance reasoning from the CoT. Further analysis and case study
 are detailed in Appendices~\ref{app:ablation_full_res}, \ref{app:further_ablation}, and \ref{app:case}.



\subsubsection{Effectiveness of MPCoT}
Table~\ref{tab:main_results} clearly illustrates the step-by-step effectiveness of our multi-perspective training pipeline.A key initial finding is the remarkable effectiveness of the base LLM, as the few-shot model achieves performance competitive with a fully-trained BERT baseline on several datasets. Building on this, a consistent performance gain is observed through the SFT and DPO stages for both single- and multi-perspective models. 

The comparison between the multi- and single-perspective approaches, however,  reveals a more critical insight: While the multi-perspective model is not always superior in the few-shot setting, it progressively improves and ultimately surpasses the best single-perspective model after the DPO stage. Our final teacher model $\text{MPCoT}_{\text{SFT+DPO}}$ consistently achieves the highest performance among all teacher models, surpassing the strongest single-perspective baseline ($\text{Best-SingleCoT}_{\text{SFT+DPO}}$) by an average of \textbf{+1.96 points} in accuracy and \textbf{+3.5 points} in F1 across all datasets. This significant improvement suggests that our SFT and DPO stages successfully train the model to resolve conflicts and adaptively select the most relevant reasoning path from multiple viewpoints, leading to more robust and accurate judgments. Notably, our approach also outperforms $\text{ProgressiveCoT}_{\text{SFT}}$, an external baseline inspired by the curriculum-based SFT strategy of \citet{Tang_2025} using progressively complex CoTs. This suggests that empowering the model to adaptively select among diverse reasoning perspectives is more effective than guiding it through a fixed curriculum of reasoning complexity.

\subsubsection{Effectiveness of LRKD}
As shown in Table \ref{tab:main_results}, all of our LRKD variants
consistently outperform the BERT baseline~\cite{kenton2019bert} in both accuracy and macro F1. In particular, the GAT-based extractor with attention mechanisms performs best, improving accuracy by \textbf{+2.37 points} and F1 by \textbf{+3.91 points} on average over the baseline, suggesting that explicit modeling of token-level interactions through graph-based message passing better captures the implicit reasoning structure embedded in LLM-generated rationales. LRKD outperforms MKD\cite{Zhao_2025} by \textbf{+1.15 points} in accuracy and \textbf{+2.49 points} in F1 because MKD trains the student to predict token-level BIO tags derived from rationales, restricting distillation to explicit and surface-level text matching, which fails to capture deeper semantic reasoning. LRKD also exceeds CED-KD \cite{agrawal-etal-2025-rationale} by \textbf{+0.98 points} in accuracy and \textbf{+2.5 points} in F1. This highlights a key limitation of training a decoder to regenerate rationales, while only using its gradient to update the encoder does not preserve much reasoning capabilities at inference time. By retaining a lightweight latent reasoning extractor in the training and inference stage, LRKD efficiently internalizes LLM reasoning in the embedding space, achieving stronger performance without text generation.

\begin{table}[htbp]
\centering
\caption{Comparison of the model efficiency.}
\label{tab:model_complexity}
\begin{tabular}{lcc}
\toprule
\textbf{Model} & \textbf{Params} & \textbf{Infer Time (ms)} \\
\midrule
\multicolumn{3}{l}{\textbf{Teacher Model:}} \\
\quad MPCoT & 14B & 46800 \\ 
\multicolumn{3}{l}{\textbf{Student Model:}} \\
BERT-multilingual-base\cite{kenton2019bert} & 168.15M & 132.22 \\
\quad $+\text{LRKD lextractor}_{\text{MLP}}$ & 169.33M & 134.02 \\
\quad $+\text{LRKD lextractor}_{\text{Poly}}$ & 168.18M & 132.68 \\
\quad $+\text{LRKD lextractor}_{\text{GAT}}$ & 168.74M & 148.76 \\
\bottomrule
\end{tabular}
\end{table}

\subsubsection{Model Complexity}
To assess practicality in real-world deployment, we analyze model complexity and inference efficiency in terms of parameter count and latency. Here, inference time (ms) represents the averaged processing time for 100 query–item pairs per batch on an NVIDIA A100 GPU. As shown in Table ~\ref{tab:model_complexity}, the results reveal that the introduction of our LRKD framework leads to only a marginal increase in both model parameters and inference latency compared to the BERT-multilingual-base baseline model. The Poly-Encoder-based variant is extremely efficient, adding merely 0.03M parameters with almost no latency increase (+0.46ms). Furthermore, The GAT-based extractor (+0.59M) incurs a higher latency cost (+16.54ms), which is a trade-off for its superior performance as shown in Table \ref{tab:main_results}.  This minimal overhead is a direct result of our design philosophy: the latent reasoning extractor is a lightweight, non-autoregressive module that operates solely in the embedding space. In contrast, the teacher LLM (MPCoT, 14B parameters) exhibits an inference time that is orders of magnitude slower (46,800 ms vs. 132 ms), rendering it infeasible for real-time applications. This highlights the core value of our distillation framework: it successfully internalizes the sophisticated reasoning capabilities of a massive LLM into a compact and efficient student model, achieving a compelling balance between performance and deployability.

\subsection{Ablation Study}
To better understand the contribution of each component in our MPCoT and LRKD framework, we conduct a series of ablation studies.  For brevity, we present results on the AliExpress and ESCI-US test sets; similar improvements are consistently observed on ESCI-JP and ESCI-ES test sets (see Appendix \ref{app:ablation_full_res} for details).  

\begin{table}[htbp]
\centering
\caption{Ablation study of multi-perspective \textit{SFT} on AliExpress and ESCI-US test sets.}
\label{tab:ablation_results_teacher}
\begin{tabular}{lcccc}
\toprule
\textbf{Model} &
\multicolumn{2}{c}{\textbf{AliExpress}} &
\multicolumn{2}{c}{\textbf{ESCI-US}} \\
\cmidrule(lr){2-3} \cmidrule(lr){4-5}
& \textbf{ACC} & \textbf{F1} & \textbf{ACC} & \textbf{F1} \\
\midrule
\textbf{User Intent SFT} & 58.49 & 51.96 & 63.20 & 50.21 \\
\quad + User Intent DPO & +4.46 & +3.98  & +5.12 & -0.14 \\
\quad $\text{+ MPCoT DPO}$ & +3.08 & +3.11  & +2.95  & -1.45 \\

\textbf{Structured Analysis SFT} & 57.81 & 51.02 & 62.63 & 48.05 \\
\quad + Structured Analysis DPO & +4.89 & +4.01 & +4.78 & +2.00 \\
\quad $\text{+ MPCoT DPO}$ & +3.44 & +2.12 & +5.79 & +0.30\\
\textbf{Business Rules SFT} & 59.81 & 53.87 & 56.89 & 45.06 \\
\quad + Business Rules DPO & +5.02 & +4.39 & +12.60 & +5.02 \\
\quad $\text{+ MPCoT DPO}$ & +3.88 & +2.01 & +9.63 & +2.56 \\
\textbf{MPCoT SFT+DPO} & \textbf{68.36} & \textbf{65.90} & \textbf{70.23} & \textbf{51.85} \\
\bottomrule
\end{tabular}
\end{table}

\begin{table}[htbp]
\centering
\caption{Ablation study of LRKD on AliExpress and ESCI-US test sets.}
\label{tab:ablation_results}
\begin{tabular}{lcccc}
\toprule
\textbf{Model} &
\multicolumn{2}{c}{\textbf{AliExpress}} &
\multicolumn{2}{c}{\textbf{ESCI-US}} \\
\cmidrule(lr){2-3} \cmidrule(lr){4-5}
& \textbf{ACC} & \textbf{F1} & \textbf{ACC} & \textbf{F1} \\
\midrule
$\textbf{LRKD}_{\text{Poly}}$ & \textbf{56.92} & \textbf{51.37} & \textbf{67.92} & \textbf{44.33} \\
\quad w/o latent reason extractor & -1.49 & -0.89 &-1.73  & -1.35 \\
\quad w/o guidance loss & -1.38 & -1.04 & -1.98 & -1.55 \\
$\textbf{LRKD}_{\text{GAT}}$ & \textbf{57.36} & \textbf{52.29} & \textbf{68.73} & \textbf{45.83} \\
\quad w/o latent reason extractor & -1.35 & -1.15  & -1.09 & -1.11 \\
\quad w/o guidance loss & -1.42  & -1.13 & -2.14 & -2.06 \\
\bottomrule
\end{tabular}
\end{table}

\subsubsection{Ablation Study on Multi-Perspective SFT}
\label{sec:ablation_sft_dpo}
We first claim that a multi-perspective approach is crucial throughout the entire SFT-to-DPO training pipeline. To validate this, we conduct an ablation in which we fine-tune single-perspective SFT models using multi-perspective DPO data ($\text{+ MPCoT DPO}$). As shown in Table~\ref{tab:ablation_results_teacher}, this approach consistently yields suboptimal results, often inferior to fine-tuning with perspective-aligned DPO data. 
This indicates that a single-perspective SFT model lacks the foundation to synthesize diverse reasoning styles, failing to leverage multi-perspective signals introduced at the DPO stage. 
Interestingly, the \textit{Business Rules} gains the most from its perspective-aligned DPO, validating DPO's effectiveness in teaching the model to leverage structured rules.

\subsubsection{Ablation Study on Perspective Diversity in DPO}
To isolate the benefit of perspective diversity from the increased number of preference pairs generated during DPO, we compare MPCoT against a size-matched single-perspective baseline. As shown in Appendix~\ref{app:dpo_with_mpcot}, MPCoT still outperforms, confirming its benefit comes from diverse reasoning perspectives rather than more data.

\subsubsection{Ablation Study on Latent Reasoning Guidance}
To assess the contribution of the latent reasoning extractor and the reasoning-guided loss, we test two key variants: (1)~\textit{w/o latent reasoning extractor}, where the latent reasoning vector is replaced with the BERT [CLS] representation $\mathbf{h}_{\text{cls}}$; and (2)~\textit{w/o reasoning guidance loss}, where the latent reasoning extractor is retained but the MSE-based alignment loss is removed. As shown in Table~\ref{tab:ablation_results}, both variants degrade performance, with the removal of the guidance loss causing a more significant drop. This indicates the extractor's performance gain stems not from more parameters, but from its ability to learn meaningful reasoning signals. Moreover, the larger drop when using only $\mathbf{h}_{\text{cls}}$ for both classification and reasoning suggests a representation overload, thereby validating the necessity of maintaining a dedicated latent reasoning extractor. We also compare LRKD against basic ensemble approaches, which combine multiple independently trained models. The results in Appendix~\ref{app:lrkd_efficiency} show our unified model achieves competitive performance while requiring only a single teacher model for distillation, thus avoiding the substantial training overhead of multi-teacher baselines.


\subsubsection{Probing for Reasoning-Specific Semantics}
\begin{figure}
\centering
\includegraphics[width=\linewidth]{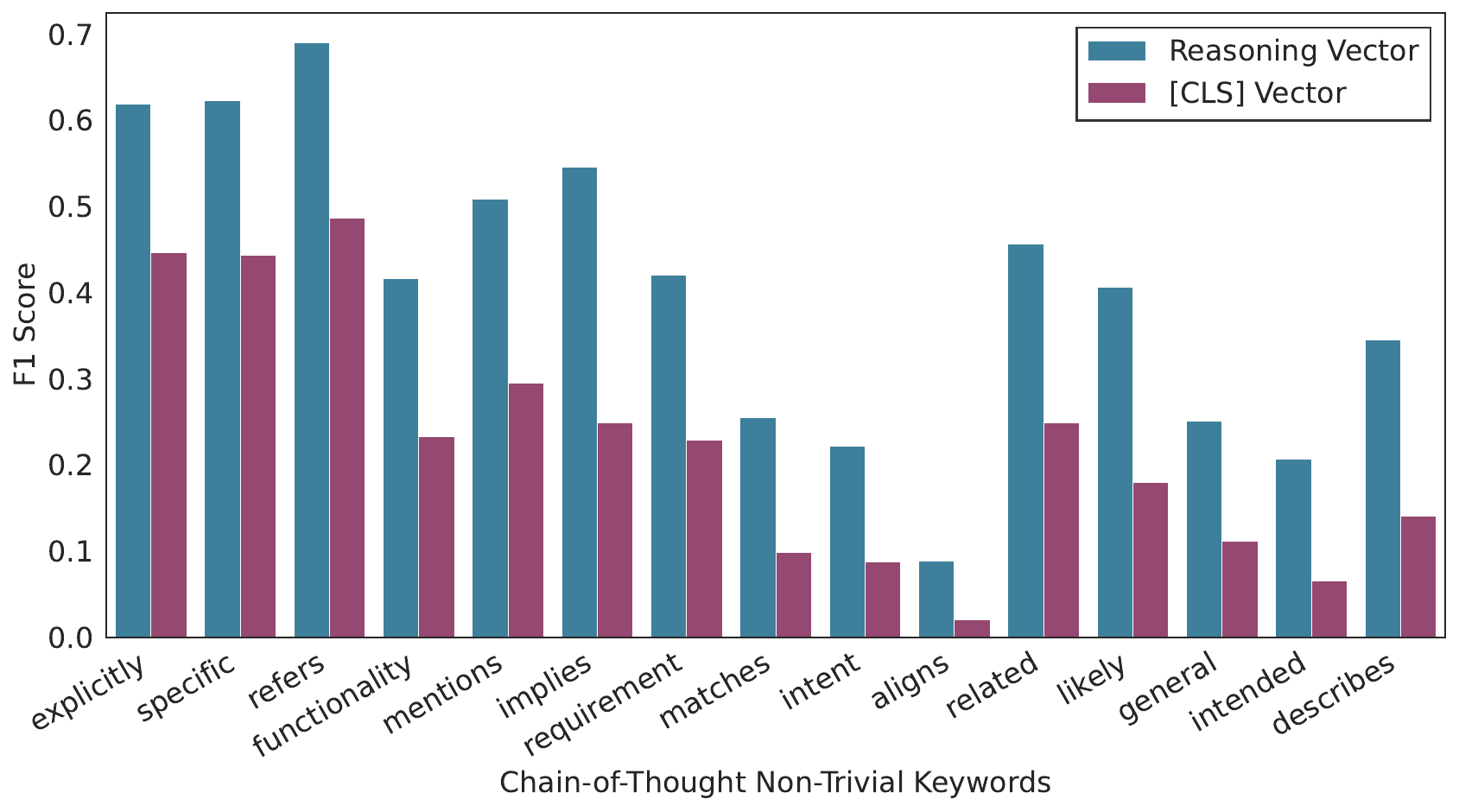} 
\caption{Probe F1 scores for top non-trivial CoT keywords}
\Description{Bar chart showing F1 scores for top non-trivial chain-of-thought keywords, comparing latent reasoning vector and [CLS] representation.}
\label{fig:probe_keyword}
\end{figure}

To verify that our latent reasoning vector captures reasoning-specific semantics rather than lexical overlaps, we design a probing task using LLM-generated CoTs on ESCI-US test dataset. We first extract the most frequent semantic keywords from CoTs and exclude any keyword appearing in the query or item title to ensure the probe cannot rely on surface matching. To assess whether our latent reasoning extractor internalizes abstract reasoning signals that are absent from the raw query–item title pair, we evaluate its ability to predict the presence of non-trivial keywords. Specifically, for each such keyword, we train a probe on either the latent reasoning vector or the [CLS] representation. As shown in Figure~\ref{fig:probe_keyword}, on the top 15 non-trivial keywords, the latent reasoning vector achieves an average F1 score of 0.403, an 81.8\% relative improvement over the [CLS] representation (0.221). This gain is statistically significant across all 15 keywords (paired t-test, $p < 0.05$ over 10 runs). Notably, the concept extractor successfully encodes abstract functional reasoning words such as "mentions", "implies", and "likely", which semantic signals are absent from the query-item title pair, but are crucial to the LLM’s relevance judgment. These results confirm that our latent reasoning extractor internalizes the intent of LLM reasoning in a structured, deployable form, going beyond surface-level lexical matching.

\subsection{Online Evaluation}
We conducted a 7-day online A/B test on AliExpress’s search advertising system, assigning 10\% of user traffic to our distilled model with latent reasoning capability and 10\% to an equivalent student model without such reasoning components. In online deployment, the relevance model outputs a relevance score for each query–item pair, which is then assigned to a discrete numeric relevance level (e.g., 0 to 4, representing \textit{Irrelevant} to \textit{Exact}) based on predefined business rules and calibrated thresholds. Query-item pairs assigned to the low relevance tiers(e.g., tier 0 or 1) are filtered before the ranking module. Our approach achieved consistent and statistically significant improvements: +1.42\% in RPM, +0.48\% in CTR and +0.4\% in human-evaluated RS score. These gains confirm that our framework delivers more intent-aligned ads, enhances user experience, improves ad exposure efficiency, and ultimately boosts commercial performance in our search advertising pipeline.

\section{Conclusion}
In this paper, we propose a novel framework that enhances e-commerce relevance modeling through Multi-Perspective Chain-of-Thought (MPCoT) and Latent Reasoning Knowledge Distillation (LRKD). First, MPCoT enables the teacher LLM model to generate diverse, complementary rationales from user intent, structured analysis, and business rule perspectives, improving both interpretability and accuracy. Second, LRKD distills this rich reasoning into a lightweight student model via a trainable latent reasoning extractor operating in embedding space, which preserves reasoning semantics without text generation at inference. Extensive offline and online experiments demonstrate consistent gains in relevance accuracy, user satisfaction, and commercial metrics. These results validate that our framework successfully bridges the gap between the sophisticated reasoning capabilities of LLMs and the stringent latency and efficiency requirements of real-world deployment, enabling systems to "think broad, act fast." We hope this work will inspire future efforts to develop more effective methods for deploying the reasoning capabilities of large language models in e-commerce, particularly in building explainable and high-performance AI systems for tasks such as product retrieval, query understanding, and personalized ranking. 


\bibliographystyle{ACM-Reference-Format}
\balance
\bibliography{bibfile}

\appendix
\section{Latent Reasoning Extractor Architectures}

\label{app:extractor_details}
Let contextualized token representations be denoted as $\mathbf{H} \in \mathbb{R}^{L \times d}$ and $M \in \{0, 1\}^L$ be the attention mask indicating valid tokens. The latent reasoning extractor is a learnable function $R$ that outputs a latent reasoning vector $r = R(\mathbf{H}, M)$. Below we detail three representative implementations used in our experiments:

\paragraph{MLP-Based Extractor.}
Each token is transformed independently via an MLP, followed by masked mean pooling:
\begin{align}
\tilde{\mathbf{H}} &= \text{MLP}(\mathbf{H}) \in \mathbb{R}^{L \times d}, \\
r &= \frac{1}{\sum_i M_i} \sum_{i=1}^L M_i \cdot \tilde{\mathbf{h}}_i.
\end{align}

\paragraph{Poly-Encoder-Based Extractor~\cite{humeaupoly}.}
We introduce $K$ learnable context codes $U = [u_1, \dots, u_K]^\top \in \mathbb{R}^{K \times d}$. Attention weights are computed as:
\begin{align}
A &= \text{softmax}\left(\frac{\mathbf{H} U^\top}{\sqrt{d}} + B\right) \in \mathbb{R}^{L \times K},
\end{align}
where $B_{ij} = 0$ if $M_i = 1$, otherwise $-\infty$. The final vector is the average over $K$ aggregated heads:
\begin{align}
C_{\text{heads}} &= A^\top \mathbf{H} \in \mathbb{R}^{K \times d}, \quad
r = \frac{1}{K} \sum_{k=1}^K c_k.
\end{align}

\paragraph{GAT-Based Extractor~\cite{velickovic2017graph}.}
We construct a graph where nodes correspond to valid tokens ($V = \{i \mid M_i = 1\}$). Edge attention scores are:
\begin{align}
\alpha_{ij} &= \text{LeakyReLU}\left(a^\top [W \mathbf{h}_i \parallel W \mathbf{h}_j]\right), \\
r_i' &= \sum_{j \in \mathcal{N}(i)} \text{softmax}_j(\alpha_{ij}) \cdot W \mathbf{h}_j,
\end{align}
where $W$ is a learnable projection and $\mathcal{N}(i)$ denotes neighbors of node $i$ (here, all valid tokens). The final reasoning vector is the mean of all node representations:
\begin{align}
r = \frac{1}{|V|} \sum_{i \in V} r_i'.
\end{align}

\begin{table*}[htbp]
\centering
\caption{Detail Ablation Results of MPCoT and LRKD.}
\label{tab:detail_ablation_results}
\begin{tabular}{lcccccccc}
\toprule
\textbf{Model} &
\multicolumn{2}{c}{\textbf{AliExpress}} &
\multicolumn{2}{c}{\textbf{ESCI-US}} &
\multicolumn{2}{c}{\textbf{ESCI-JP}} &
\multicolumn{2}{c}{\textbf{ESCI-ES}} \\
\cmidrule(lr){2-3} \cmidrule(lr){4-5}\cmidrule(lr){6-7} \cmidrule(lr){8-9}
& \textbf{ACC} & \textbf{F1} & \textbf{ACC} & \textbf{F1} & \textbf{ACC} & \textbf{F1} & \textbf{ACC} & \textbf{F1} \\
\midrule
\textbf{User Intent SFT } & \textbf{58.49} & \textbf{51.96} & \textbf{63.20} & \textbf{50.21} & \textbf{60.11} & \textbf{49.97}  & \textbf{60.62} & \textbf{55.28} \\
\quad + User Intent DPO & +4.46 & +3.98  & +5.12 & -0.14  & +5.28 & +1.17 & +7.48 & +2.93 \\
\quad $\text{+ MPCoT DPO}$ & +3.08 & +3.11  & +2.95  & -1.45 & +3.63 & +0.14 & +4.79 & -1.29\\

\textbf{Structured Analysis SFT} & \textbf{57.81} & \textbf{51.02} & \textbf{62.63} & \textbf{48.05} & \textbf{59.67} & \textbf{47.61} & \textbf{59.95} & \textbf{51.48} \\
\quad + Structured Analysis DPO & +4.89 & +4.01 & +4.78 & +2.00& +6.45 & +1.45 & +7.48 & +6.19 \\
\quad $\text{+ MPCoT DPO}$ & +3.44 & +2.12 & +5.79 & +0.30 & +3.76 & -0.19 & +2.92 & +0.06 \\
\textbf{Business Rules SFT} & \textbf{59.81} & \textbf{53.87} &\textbf{ 56.89} & \textbf{45.06 } & \textbf{55.53} & \textbf{46.90} & \textbf{55.07} & \textbf{49.92} \\
\quad + Business Rules DPO & +5.02 & +4.39 & +12.60 & +5.02 & +11.35 & +2.08 & +12.27 & +7.63\\
\quad $\text{+ MPCoT DPO}$ & +3.88 & +2.01 & +9.63 & +2.56 & +8.59 & +1.02 & +7.94 & +3.22\\
\textbf{MPCoT SFT+DPO} & \textbf{68.36} & \textbf{65.90} & \textbf{70.23} & \textbf{51.85}& \textbf{69.72} & \textbf{53.69} & \textbf{67.32} & \textbf{60.31} \\
\addlinespace
$\textbf{LRKD}_{\text{Poly}}$ & \textbf{56.92} & \textbf{51.37} & \textbf{67.92} & \textbf{44.33}  & \textbf{63.44} &  \textbf{45.27} & \textbf{65.89} & \textbf{55.47} \\
\quad w/o latent reason extractor & -1.49 & -0.89 &-1.73  & -1.35 & -2.34& -0.98 & -1.92& -1.40\\
\quad w/o guidance loss & -1.38 & -1.04 & -1.98 & -1.55 & -2.39 & -1.03 & -2.95 & -1.51 \\
$\textbf{LRKD}_{\text{GAT}}$ & \textbf{57.36} & \textbf{52.29} & \textbf{68.73} & \textbf{45.83} & \textbf{63.96}  & \textbf{47.21} & \textbf{66.30} & \textbf{55.89} \\
\quad w/o latent reason extractor & -1.35 & -1.15  & -1.09 & -1.11 & -1.76 & -0.95 & -2.41 & -1.22  \\
\quad w/o guidance loss & -1.42  & -1.13 & -2.14 & -2.06 & -3.03 & -1.62 & -2.88 & -1.34 \\
\bottomrule
\end{tabular}
\end{table*}

\begin{table}[htbp!]
\centering
\caption{Ablation studies on perspective diversity and architectural efficiency.}
\label{tab:ablation_mpcot_lrkd}
\begin{tabular}{lcccc}
\toprule
\textbf{Model} &
\multicolumn{2}{c}{\textbf{AliExpress}} &
\multicolumn{2}{c}{\textbf{ESCI-US}} \\
\cmidrule(lr){2-3} \cmidrule(lr){4-5}
& \textbf{ACC} & \textbf{F1} & \textbf{ACC} & \textbf{F1} \\
\midrule
\textbf{Teacher} & & & & \\
\quad $\text{S-CoT}_{\text{SFT+DPO}}$ (size-match) & 66.49 & 62.50 & 69.98 & 50.05 \\
\quad $\text{MPCoT}_{\text{SFT+DPO}}$  & \textbf{68.36} & \textbf{65.91} & \textbf{70.23} & \textbf{51.85} \\
\midrule
\textbf{Student} & & & & \\
\quad Avg. Ensemble & 55.10 & 50.01 & 68.05 & 42.33 \\
\quad LRKD-Combined & 57.28 & 51.86 & \textbf{68.76} & 45.02 \\
\quad \textbf{LRKD-GAT (Ours)} & \textbf{57.36} & \textbf{52.29} & 68.73 & \textbf{45.83} \\
\bottomrule
\end{tabular}
\end{table}

\section{Implementation Details}
\label{app:implementation_details}
\paragraph{LLM-based Teacher Model Training Setup}
All teacher model variants are built upon Qwen3-14B~\cite{yang2025qwen3} as the shared backbone architecture to ensure a fair comparison. For the initial few-shot CoT generation, we sample 5 rationales for each input using a temperature of 0.7, a top-p of 0.99, and a top-k of 50. During the SFT stage, we employ LoRA for parameter-efficient fine-tuning. The model is trained for 3.0 epochs with a global batch size of 32 (8 per device across 4 A100 GPUs), a maximum sequence length of 3072, and a learning rate of 5e-5, managed by a cosine scheduler. In the subsequent DPO stage, training is conducted for 3.0 epochs with an effective batch size of 64 (achieved with a per-device batch size of 1 and 16 gradient accumulation steps across 4 NVIDIA A100 GPUs). The learning rate is set to 5.0e-6 with a cosine scheduler and a warmup ratio of 0.1, and the training is performed using bf16 precision.

\paragraph{Student Model Knowledge Distillation Setup}
For fair comparison, we conducted offline experiments using BERT-multilingual-base\cite{kenton2019bert} as the shared backbone architecture across all variants in the classification task. In LRKD, the LLM-generated CoT is encoded into a fixed semantic target using the frozen BGE-M3\cite{chen2024bgem3embeddingmultilingualmultifunctionality} sentence encoder with max sequence length 1024, while the input query–item title pair is truncated to a maximum length of 128 tokens. The student model is optimized using the AdamW optimizer with a learning rate of 2e-6 and trained for 5 epochs with 256 batch size on NVIDIA A100 GPUs. The weighting coefficient $\lambda$ for the loss of latent reasoning guidance is set to 0.1. For the extractor architectures, we set the hidden dimension to 128 across the board: the MLP uses 2 layers, the GAT uses 1 layer, and the Poly-Encoder employs 32 learnable context codes. The rival methods MKD~\cite{Zhao_2025} and CED-KD~\cite{agrawal-etal-2025-rationale}, are implemented following the settings described in their respective papers.

\section{Detail Ablation Results of MPCoT and LRKD}
\label{app:ablation_full_res}
Table~\ref{tab:detail_ablation_results} details the performance of all model variants for both the teacher (MPCoT) and student (LRKD) frameworks across all evaluation datasets.

\section{Further Ablation Results}
\label{app:further_ablation}
\subsection{Effect of Perspective Diversity in DPO}
\label{app:dpo_with_mpcot}

In our main results (Table~\ref{tab:main_results}), our final model $\text{MPCoT}_{\text{SFT+DPO}}$ consistently outperforms its best single-perspective counterpart. Since MPCoT constructs preference pairs by comparing correct and incorrect rationales across different perspectives, it naturally yields more training examples than single-perspective approaches. A critical question is whether this gain stems from the richer cross-perspective reasoning signals or simply from the larger DPO dataset that MPCoT inherently produces. To investigate this, we trained a single-perspective teacher on a dataset expanded to match the size of our MPCoT training set, denoted as $\text{S-CoT}_{\text{SFT+DPO}}$. 

The results in Table~\ref{tab:ablation_mpcot_lrkd} show that our MPCoT teacher outperforms this size-matched baseline. This confirms that the performance gains are attributable to the diverse reasoning traces provided by our framework, not simply data quantity. This finding, coupled with the observation from Section~\ref{sec:ablation_sft_dpo} that a multi-perspective foundation is crucial from the SFT stage, validates that the synergy between diverse SFT and DPO stages is the key driver of our model's superior performance.

\subsection{Architectural Efficiency of LRKD}
\label{app:lrkd_efficiency}

To evaluate the architectural design of LRKD, we compare it against two straightforward alternatives that also aim to incorporate multi-perspective signals:
\begin{enumerate}
    \item \textbf{Avg. Ensemble}: A simple baseline that trains three separate BERT models (one per perspective) and averages their final predictions  at inference.
    \item \textbf{LRKD-Combined}: A variant that applies LRKD to a fused signal from three independent single-perspective teachers. Specifically, if $r_1, r_2, r_3$ are the CoTs from the three teachers, they are concatenated with a separator token encoded into a unified embedding vector: 
    $$ \mathbf{e}_{\text{combined}} = \text{BGE}(r_1, r_2, r_3) \in \mathbb{R}^d $$
\end{enumerate}

As shown in Table~\ref{tab:ablation_mpcot_lrkd}, our unified LRKD-GAT model outperforms the Avg. Ensemble despite using only a single student and a single multi-perspective teacher. Although LRKD-Combined achieves competitive results, it incurs approximately three times the offline teacher-training cost, highlighting the efficiency of our unified multi-perspective distillation framework. This tradeoff starkly highlights the superior efficiency of our integrated MPCoT-to-LRKD pipeline, which delivers strong performance without resorting to costly multi-teacher or multi-model architectures.

\section{Case Study}
\label{app:case}

\begin{figure*}[htbp]
    \centering
    \begin{minipage}{1.0\textwidth}
        \centering
        \includegraphics[width=\linewidth]{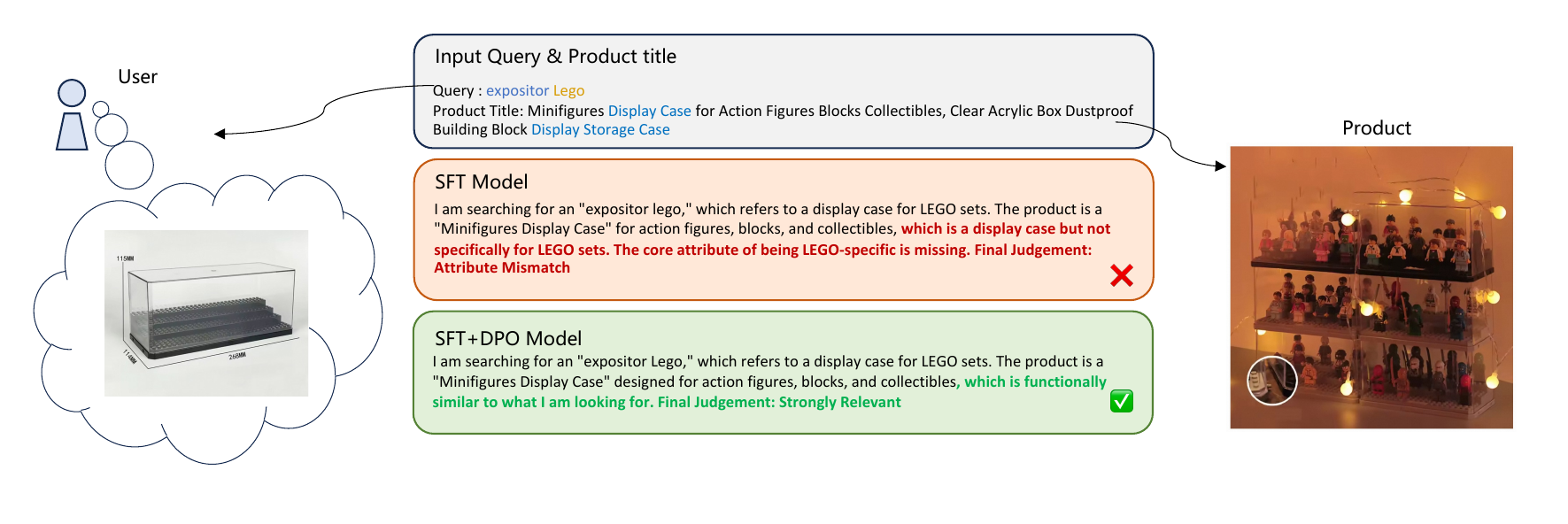}
        \centerline{(a) Case study of the DPO process.}
    \end{minipage}
    \hfill 
    \begin{minipage}{1.0\textwidth}
        \centering
        \includegraphics[width=\linewidth]{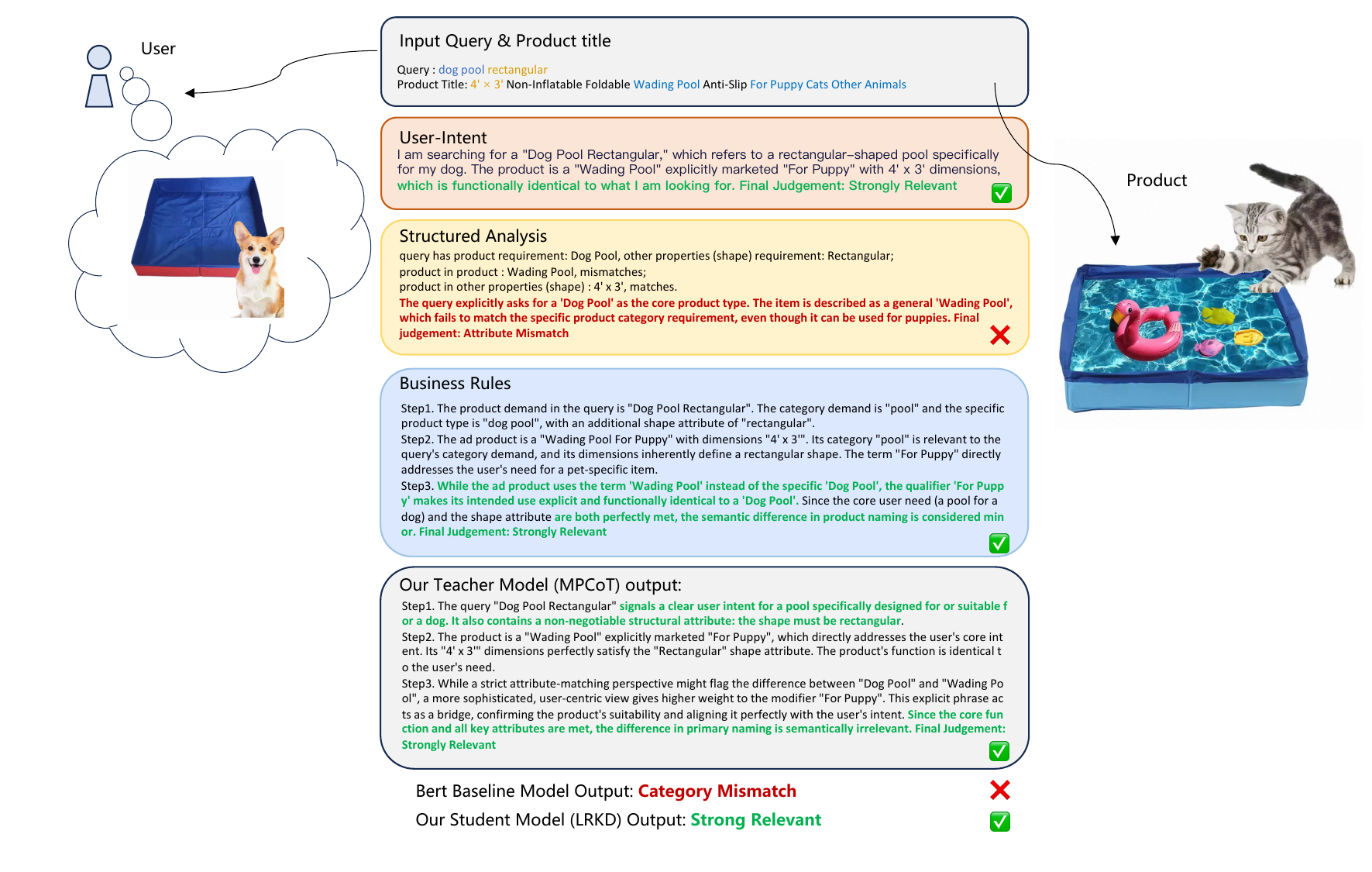}
        \centerline{(b) Case study of the multi-perspective CoT.}
    \end{minipage}
    \caption{Case studies from the AliExpress Dataset.}
    \label{fig:case_1}
\end{figure*}

To provide qualitative insights into our framework, we present two case studies in Figure~\ref{fig:case_1}. First, we illustrate the importance of the DPO process in refining the model’s understanding of domain-specific nuances, such as functional substitution, as shown in Figure~\ref{fig:case_1}(a). For the query “expositor Lego,” the SFT-only model adheres to strict semantic matching and incorrectly classifies the product as \textit{Attribute Mismatch} because the “Lego” brand name is absent. In contrast, after the DPO phase, the model learns to prioritize functional intent over literal keyword matching. It correctly identifies the product as \textit{Strongly Relevant} by recognizing that it is “functionally similar” to the user’s need, demonstrating that DPO is essential for endowing the model with this real-world e-commerce reasoning capability.

Next, Figure~\ref{fig:case_1}(b) illustrates the effectiveness of the full MPCoT and LRKD framework on a challenging example. The task involves a query for a "dog pool rectangular" and an ad titled "Wading Pool...For Puppy." A standard BERT model fails to recognize their functional similarity and misclassifies the pair as \textit{Category Mismatch}. This difficulty is further exacerbated by conflicting signals from single-perspective reasoning: the rigid \textit{Structured Analysis} perspective fails, whereas the more flexible \textit{User Intent} and \textit{Business Rules} perspectives correctly identify relevance. Our MPCoT teacher resolves this conflict by synthesizing multiple viewpoints and prioritizing functional intent over literal naming discrepancies, yielding a correct \textit{Strongly Relevant} judgment. This correct reasoning is then successfully distilled into our LRKD student model, which mirrors the teacher's conclusion, demonstrating the effectiveness of our end-to-end framework.


\end{document}